\documentclass[aps,prl,showpacs]{revtex4}
\usepackage{graphicx}

\begin{document}

\title{High photon number path entanglement in the
interference of spontaneously downconverted photon pairs with coherent laser light}


\author{Holger F. Hofmann}
\email{h.hofmann@osa.org}
\affiliation{
Graduate School of Advanced Sciences of Matter, Hiroshima University,
Kagamiyama 1-3-1, Higashi Hiroshima 739-8530, Japan}

\author{Takafumi Ono}
\affiliation{
Graduate School of Advanced Sciences of Matter, Hiroshima University,
Kagamiyama 1-3-1, Higashi Hiroshima 739-8530, Japan}

\begin{abstract}
We show that the quantum interference between downconverted photon
pairs and photons from coherent laser light can produce a maximally path
entangled $N$-photon output component with a fidelity greater than 90\%
for arbitrarily high photon numbers. A simple beam splitter operation can
thus transform the 2-photon coherence of down-converted light into an almost
optimal $N$-photon coherence.
\end{abstract}

\pacs{
42.50.Dv 
03.67.Mn 
03.65.Ud 
42.50.Ar 
}

\maketitle


One of the most challenging tasks of optical quantum information
technologies is the generation and maintenance of entanglement
in multi-photon states. Perhaps the most extreme case of $N$-photon
entanglement is the path entangled state, where $N$ photons are in
a superposition of being either all in one optical path or all in the
other optical path inside a two path interferometer. In terms of the
photon number states inside the interferometer, this state can be
written as a superposition of two maximally distinguishable $N$-photon
states,
\begin{equation}
\label{eq:noon}
\mid \! \mbox{NOON} \rangle = \frac{1}{\sqrt{2}}
(\mid \! N;0 \rangle + \mid \! 0;N \rangle).
\end{equation}
Since path entangled states represent a pure $N$-photon coherence,
they are optimally sensitive to small phase shifts between the two
optical paths. The generation of path entangled states with high photon
number could therefore improve the precision of phase measurements to the
fundamental quantum mechanical limit \cite{Ou97,Gio04,Bou04}.
Various methods for generating path entangled states with
arbitrary numbers of photons have been proposed
\cite{Kok02,Fiu02,Pry03,Hof04}, but up to now,
all known methods become more and more complicated as
photon number increases, making it difficult to implement them for
more than the three or four photon states already realized
experimentally \cite{Wal04,Mit04}.
Specifically, these experimental
demonstrations were based on post-selection of the output,
which means that only a very limited fraction of the actual $N$-photon
emissions of the sources could contribute to the observed $N$-photon interferences. It is therefore of great interest to develop sources
that provide $N$-photon path entanglement at high fidelity without
requiring any post-selection of the output.

In the case of three photons, a particularly simple and elegant way
to generate path entanglement is by an interference between coherent
light and downconverted photon pairs at a beam splitter \cite{Sha04}.
This method can be interpreted as a cancellation of all output components
other than $\mid \! N; 0 \rangle$ and $\mid \! 0; N \rangle$ by destructive
quantum interference. Unfortunately, the exact cancellation of unwanted
terms cannot be achieved at photon numbers higher than three, so that
the extension to higher photon numbers appears to be difficult \cite{Liu06}.
Nevertheless, it may be interesting to consider just how close we can get
to ideal path entangled $N$-photon states by using only the basic two
mode interference of downconverted photon pairs with laser light at
a single beam splitter.

In this paper, we show that the interference between coherent laser light
and downconverted photon pairs at a beam splitter can produce path
entangled states of arbitrarily high photon number $N$ with fidelities
greater than 90\%. In fact, the fidelity of the path entanglement increases
with larger photon numbers $N$, approaching a maximum of $\sqrt{8/9}$ or
94.3\% at very high $N$.
The reason for this surprisingly strong non-classical effect is the
combination of a perfect two photon coherence in the downconverted
light with the lower photon number fluctuations of the coherent
state of the laser. When the average photon numbers contributed by
each field are nearly equal, this results in a slightly anti-squeezed photon
number distribution around the maximum at $\mid \! N/2,N/2 \rangle$,
with a perfect multi-photon interference pattern ``imprinted''
upon it. By sending this multi-photon interference pattern ``back''
through a beam splitter, the multi-photon coherence is recovered and
path entanglement results.
The $N$-photon coherence observed in the interference between
coherent laser light and downconverted photon pairs can thus be
understood as a reversal of the interference process that is normally
used to verify path entanglement.



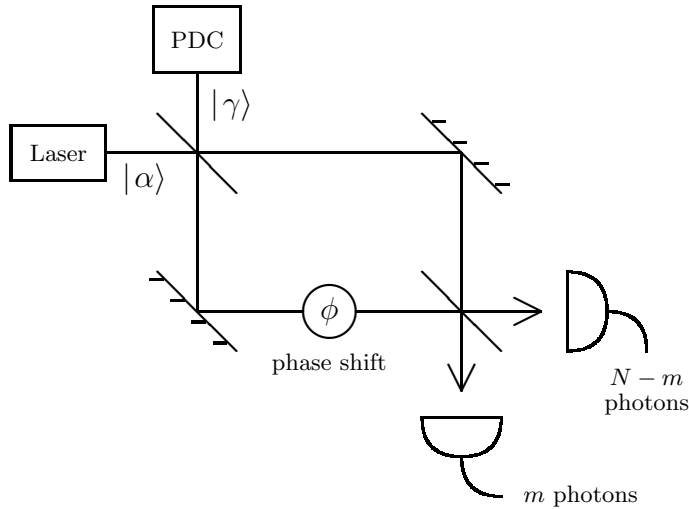
\begin{figure}
\begin{picture}(300,210)
\thicklines
\put(84,171){\framebox(32,24){\small PDC}}
\put(103,148){\makebox(20,20){\large $\mid \! \gamma \rangle$}}
\put(100,170){\line(0,-1){90}}
\put(30,130){\framebox(35,20){Laser}}
\put(70,120){\makebox(20,20){\large $\mid \! \alpha \rangle$}}
\put(65,140){\line(1,0){135}}
\put(85,155){\line(1,-1){30}}
\put(185,95){\line(1,-1){30}}
\put(185,155){\line(1,-1){30}}
\put(189,152){\line(1,0){5}}
\put(197,144){\line(1,0){5}}
\put(205,136){\line(1,0){5}}
\put(213,128){\line(1,0){5}}
\put(85,95){\line(1,-1){30}}
\put(87,92){\line(-1,0){5}}
\put(95,84){\line(-1,0){5}}
\put(103,76){\line(-1,0){5}}
\put(111,68){\line(-1,0){5}}
\put(100,80){\line(1,0){40}}
\put(150,80){\circle{20}}
\put(142,72){\makebox(16,16){\large $\phi$}}
\put(135,55){\makebox(30,10){phase shift}}
\put(160,80){\line(1,0){70}}
\put(230,80){\line(-2,1){10}}
\put(230,80){\line(-2,-1){10}}
\put(200,140){\line(0,-1){90}}
\put(200,50){\line(1,2){5}}
\put(200,50){\line(-1,2){5}}

\put(185,40){\line(1,0){30}}
\bezier{100}(185,40)(185,25)(200,25)
\bezier{100}(200,25)(215,25)(215,40)
\bezier{100}(200,25)(200,10)(215,10)

\put(240,65){\line(0,1){30}}
\bezier{100}(240,65)(255,65)(255,80)
\bezier{100}(255,80)(255,95)(240,95)
\bezier{100}(255,80)(270,80)(270,65)

\put(235,50){\makebox(70,10){$N-m$}}
\put(235,40){\makebox(70,10){photons}}
\put(210,5){\makebox(70,10){$m$ photons}}

\end{picture}

\caption{\label{setup}
Schematic setup for the observation of $N$-photon path entanglement in the
interference between coherent laser light and photon pairs generated by parametric downconversion (PDC) using a Mach-Zehnder interferometer with a variable phase shift $\phi$. $N$-photon interference fringes can be observed
in the output probabilities of the coincidences between $N-m$ photons
in one output port and $m$ photons in the other output port.}
\end{figure}

A possible experimental setup for the observation of path entanglement in
the interference of laser light and downconverted photon pairs is shown
in fig.\ref{setup}. The quantum state of the input light is represented
by a coherent state $\mid \! \alpha \rangle$ describing the
laser light in input mode $\hat{a}$ and a squeezed vacuum state
$\mid \! \gamma \rangle$ describing the downconverted photon pairs in
input mode $\hat{b}$. The light interferes at the input beam splitter
and passes through the two arms of a Mach-Zehnder interferometer.
In the output, the
$N$-photon coherence between the arms of the interferometer is
observed in the phase dependence of the $N$-photon distribution between
the two output modes.
The key point about the $N$-photon interferences
observed in the output is that it is not possible to distinguish the
photons emitted by the laser from the photons generated by
downconversion \cite{Lu02}.
Therefore, each $N$-photon interference pattern is defined by a superposition
of components $\mid \! N\!-\!2k; 2k \rangle$ describing the emission of $N\!-\!2k$ photons by the laser and the generation of $k$ photon pairs by parametric downconversion.


The coherent amplitudes of the components originate from the
local coherences between the photon number states in each input mode.
In the case of the coherent state $\mid \! \alpha \rangle$ describing the
laser light in input mode $\hat{a}$ and the squeezed vacuum state
$\mid \! \gamma \rangle$ describing the downconverted photon pairs in
input mode $\hat{b}$, these local coherences can be represented by the
effects of the respective annihilation operator on each state,
\begin{eqnarray}
\hat{a} \mid \! \alpha \rangle &=& \alpha \; \mid \! \alpha \rangle,
\nonumber \\
\hat{b} \mid \! \gamma \rangle &=& \gamma \, \hat{b}^\dagger \! \mid \! \gamma \rangle.
\end{eqnarray}
The quantum coherence of the normalized $N$-photon component
$\mid \! \eta \rangle_N$ of the two mode light field can then be represented
by combining the local coherences in such a way that the operators
preserve the total photon number. A convenient representation of the
$N$-photon coherence then reads
\begin{equation}
\label{eq:class}
\hat{a}^\dagger \hat{b} \mid \! \eta \rangle_N =
\eta \frac{1}{N}(\hat{a}^\dagger \hat{a}) \hat{b}^\dagger \hat{a}
\mid \! \eta \rangle_N, \hspace{0.3cm}\mbox{where} \hspace{0.3cm}
\eta=\frac{N \gamma}{\alpha^2}.
\end{equation}
Significantly, the normalized $N$-photon state generated by the interference
of laser light and downconverted light depends only on the ratio
between the squeezing amplitude $\gamma$ and the squared coherent state
amplitude $\alpha$. It is therefore possible to generate all of the
interferometric states defined by eq. (\ref{eq:class}) with an arbitrarily
fixed downconversion amplitude $\gamma$ by adjusting only the
amplitude of the coherent laser light.


Eq.(\ref{eq:class}) uniquely defines the quantum state $\mid \! \eta \rangle_N$
and it is in principle no problem to derive the exact expansion into
the input photon number basis $\mid \! N\!-\!2k; 2k \rangle$. However,
it is easier to understand the multi-photon coherence of the state by
looking at the approximate solution in the limit of high photon numbers.
It is then possible to simplify the effects of the annihilation and
creation operators by using the approximation
$\sqrt{n+1}\approx \sqrt{n}$. The relation between the amplitudes of the
photon number components of $\mid \! \eta \rangle_N$ can then be
approximated by
\begin{equation}
\langle N\!-\!2(k\!+\!1); 2(k\!+\!1 ) \mid  \eta \rangle_{N\to\infty}
\approx \eta \left(1-\frac{2 k}{N}\right)
\langle N\!-\!2k; 2k  \mid  \eta \rangle_{N\to\infty}.
\end{equation}
By subtracting the amplitude of $k$ from the amplitude of $k+1$,
the $k$ dependence of the amplitudes can be expressed in terms
of a differential equation,
\begin{equation}
\frac{d}{dk} \langle N\!-\!2k; 2k  \mid  \eta \rangle_{N\to\infty}
\approx - \frac{\eta}{N} \left( 2k - N (1-\frac{1}{\eta}) \right)
\langle N\!-\!2k; 2k  \mid  \eta \rangle_{N\to\infty}.
\end{equation}
This differential equation is solved by a Gaussian centered
around $2k = N (1-1/\eta)$, with a variance of $N/(2\eta)$ in $k$.
At high photon numbers, the interferometric states can therefore
be approximated by Gaussian two photon coherent states. For
$\eta=2$, the maximal amplitude of the Gaussian state is located at
$\mid \! N/2 ; N/2 \rangle$, indicating equal numbers of photons
contributed by each input mode,
\begin{equation}
\label{eq:approx}
\mid \! \eta=2 \rangle_{N\to\infty} \approx
\left(\frac{4}{\pi N}
 \right)^{1/4} \; \sum_{k=0}^{N/2} \exp \left(-\frac{(2k-N/2)^2}{2 N}\right)
\mid \! N\!-\!2k; 2k \rangle.
\end{equation}

To compare this interferometric state with the path entangled state
given by eq.(\ref{eq:noon}), we have to consider what kind of
input state would result in ideal path entanglement inside the
interferometer. We can do this by applying a beam splitter transformation
to the photon number basis used in eq.(\ref{eq:noon}).
When the $\mid \! N ; 0 \rangle$ component of this state is transformed by
a beam splitter, this results in a binomial distribution of the
photons between the two modes, with a maximal amplitude
at $\mid \! N/2 ; N/2 \rangle$. Quantum interference with the
$\mid \! 0 ; N \rangle$ component eliminates the components
with an odd photon number in one of the modes. At high photon
number, the path entangled state can therefore also be approximated
by a Gaussian two photon coherent state centered at
$\mid \! N/2 ; N/2 \rangle$,
\begin{equation}
\label{eq:highNOON}
\mid \! \mbox{NOON} \rangle_{N\to\infty} \approx
\left(\frac{8}{\pi N}
 \right)^{1/4} \; \sum_{k=0}^{N/2} \exp \left(-\frac{(2k-N/2)^2}{N}\right)
\mid \! N\!-\!2k; 2k \rangle
\end{equation}

In fact, the only difference
between the $\eta=2$ interferometric state given in eq.(\ref{eq:approx}) and
the path entangled state in eq.(\ref{eq:highNOON}) is the variance of the Gaussian,
which is twice as high for the $\eta=2$ state. In the limit of high
photon number, the overlap
between the maximally path entangled state and the $N$-photon component
of the interference between coherent light and downconversion is thus
equal to the overlap of two Gaussian states with a variance ratio of two,
\begin{equation}
\label{eq:fidelity}
\left|\langle \mbox{NOON} \! \mid \! \eta=2 \rangle_{N\to\infty}
\right|^2
\approx \sqrt{\frac{8}{9}}.
\end{equation}
At high photon numbers, we can therefore expect to find a maximally
path entangled state with a fidelity approaching
$\sqrt{8/9}\approx 0.943$, or $94.3\%$.


Due to the high fidelity of the maximally path entangled state,
the interference between laser light and downconverted photon pairs
can produce high visibility $N$-photon interference fringes
without any post-selection. This is a significant difference that
distinguishes the present method of obtaining $N$-photon interference
from previous experimental realizations
\cite{Wal04,Mit04,Sun06,Res07,Nag07}. To the best of our knowledge,
the present method of obtaining $N$-photon interference fringes is
the first method that can overcome the phase super sensitivity
threshold discussed in \cite{Res07,Nag07} at arbitrarily high photon
number $N$ using only standard quantum optical technologies.
A quick and efficient estimate of the $N$-photon visibility achieved
by the $\eta=2$ state can be obtained by noting that the $\eta$ states
are all orthogonal to the path entangled state with the opposite phase,
$(\mid \! N ; 0 \rangle-\mid \! 0 ; N \rangle)/\sqrt{2}$. Therefore,
the coherence between the $\mid \! N;0 \rangle$
and the $\mid \! 0;N \rangle$ components in the photon number basis
of the paths inside the interferometer is equal to half the fidelity,
and the visibility of the $N$-photon fringes will be equal
to the fidelity. Since the phase sensitivity of an $N$ photon fringe
with visibility $V$ is given by $1/(\delta \phi) = V N$, we can expect
a phase sensitivity of $0.94 N$ in the high photon number limit, a
result that is only slightly lower than the Heisenberg limit of $N$
achieved by maximal path entanglement.
It should also be noted that experimental errors presently limit
multi-photon visibilities to much lower values. It is therefore
likely that the small deviation of the $\eta=2$ state from the
maximally path entangled state will not be the major factor limiting
the phase sensitivity achieved in an actual experiment.


\begin{figure}
\begin{picture}(300,200)
\put(115,170){\makebox(100,20){$|\langle \mbox{NOON} \mid
 \eta=2 \rangle_N|^2$}}
\put(25,20){\makebox(250,160){\scalebox{1}[1]{\includegraphics{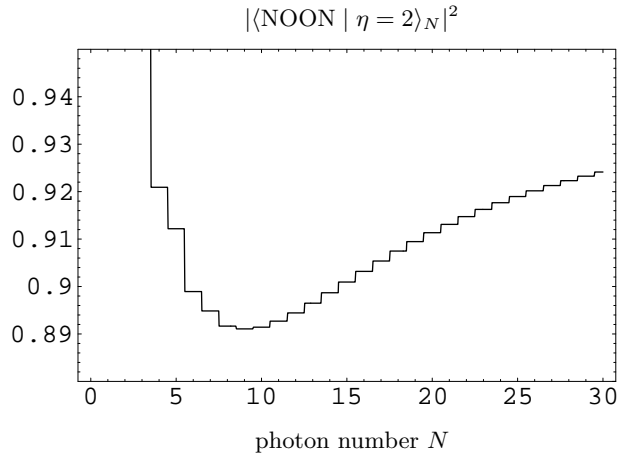}}}}
\put(115,10){\makebox(100,20){photon number $N$}}
\end{picture}
\caption{\label{fidelity}
Overlap between the $\eta=2$ interferometric state $\mid \eta=2 \rangle_N$
and the ideal path entangled state $\mid \mbox{NOON} \rangle_N$ as a
function of total photon number $N$.
}
\end{figure}

To confirm the validity of our approximations and their application to
photon numbers as low as the experimentally realized four photon
interferences, we have analyzed the precise overlap between the $\eta=2$
states and the corresponding path entangled states. This numerical
analysis is based on the exact analytical forms of the quantum states given by
\begin{eqnarray}
\label{eq:exact}
\mid \! \eta \rangle_N &=& C_{\mbox{\small norm}} \sum_{k=0}^{N/2}
\frac{1}{k!}\sqrt{\frac{(2k)! N!}{(N-2k)!}}
\left(\frac{\gamma}{2 \alpha^2}\right)^k \mid \! N\!-\!2k; 2k \rangle,
\\
\label{eq:exactNOON}
\mid \! \mbox{NOON} \rangle_N &=&
\sum_{k=0}^{N/2} \sqrt{\frac{N!}{(N-2k)! (2k)!}}
\left(\frac{1}{\sqrt{2}}\right)^{N-1} \mid \! N\!-\!2k; 2k \rangle,
\end{eqnarray}
where $C_{\mbox{\small norm}}$ is the normalization factor of the
interferometric state. Fig. \ref{fidelity} shows the overlap of the
$\eta=2$ state with the
maximally path entangled state as a function of total photon number $N$. Interestingly, the overlap has a minimum of $89.1\%$ at $N=9$ photons.
While this is already a rather high fidelity, it is possible to improve it
somewhat by varying $\eta$.
The results of this optimization for $N=2$ to $N=15$ and for $N=100$
are summarized in table \ref{optimal}. These results indicate that
fidelities of $92\%$ or greater can be obtained at any photon number
$N$.

\begin{table}
\caption{\label{optimal}
Optimized fidelities $F=|\langle \mbox{NOON} \mid \eta \rangle|^2$ for
photon numbers ranging from $N=2$ to $N=15$. The result for $N=100$ is
included as an illustration of the high photon number limit.}
\vspace{0.2cm}
\begin{tabular}{ccc|ccccccccccccccccccccccccccccccc|}
\hline \hline
& $N$ &&& 2 && 3 && 4 && 5 && 6 && 7 && 8 && 9 && 10 && 11 && 12 && 13 && 14 && 15 && 100 &
\\ \hline
& $\eta$ &&& 2.00 && 3.00 && 2.31 && 2.48 && 2.36 && 2.36 && 2.32 && 2.30 && 2.28 && 2.26 && 2.24 &&
2.22 && 2.21 && 2.19 && 2.02 &
\\[0.1cm]
& $F$ &&& 100\% && 100\% && 93.3\% && 94.1\% && 92.4\% && 92.4\% && 92.0\% && 92.0\% && 92.0\%
&& 92.0\% && 92.1\% && 92.1\% && 92.2\% && 92.3\% &&
94.1\% &
\\[0.1cm]
\hline \hline
\end{tabular}
\vspace{0.5cm}
\end{table}



Experimentally, the path entanglement of the $\eta=2$ interferometric
state can be observed by conventional $N$-photon coincidence measurements,
where the contribution of components greater than $N$ can be neglected
if the amplitudes are sufficiently small. Since no post-selection is
necessary, all $N$-photon emission events provide a valid measurement
outcome, so the detection efficiency is limited only by the available
photon counting technologies. The limitation of fidelity to between
$92\%$ and $94.3\%$ should be a small price to pay for the resulting
intrinsic efficiency of $100\%$ for the interferometric method.
In fact, previous experimental realizations of maximally path entangled
states had much lower fidelities (as indicated e.g. by the raw data
visibilities of 61\% in \cite{Wal04} and 42\% in \cite{Mit04}).
It is therefore likely that the simple and efficient method of obtaining
path entanglement presented here will outperform methods that could
achieve maximal path entanglement if experimental conditions were ideal.

In practice, present downconversion methods are unfortunately still
limited to rather low photon numbers. It will therefore be extremely
difficult to observe coincidences of more than eight photons with a
conventional setup. However, the validity of our results
at arbitrarily high photon numbers indicates that $N$-photon path
entanglement is in principle available whenever a squeezed vacuum
interferes with laser light of an appropriate amplitude. If the
technical problem of realizing precise photon number resolving
measurements can be solved, it will therefore be possible to
realize sources of path entanglement with intensities far
greater than those presently feasible.

It may be worth noting that our analysis not only provides a new and
more efficient method of generating path entanglement, it also
illustrates an interesting fundamental relation between the two photon
coherence observed in spontaneous parametric downconversion and vacuum
squeezing and the $N$-photon coherence of path entanglement. Specifically,
the interference pattern that is a typical feature of path entanglement is in fact a two photon coherence, as shown by the photon number expansions
of the ideal path entangled states given by eqs. (\ref{eq:highNOON})
and (\ref{eq:exactNOON}). Two mode interference
at a beam splitter thus transforms $N$-photon path entanglement into two
photon coherence and vice versa. This relation between two photon
coherence and entanglement in optical quantum systems may have a
a broader significance for the characterization of optical entanglement
in terms of higher order coherences \cite{Aga05,Shc05,Hil06}.


In conclusion, our results clearly show that $N$-photon path entanglement
occurs naturally in the interference between coherent laser light and
downconverted photon pairs. The high phase sensitivity of multi-photon
path entanglement can therefore be achieved without any artificial tailoring
of multi-photon interferences by post-selection or additional non-linear
processes. Instead, it is sufficient to interfer the photon pairs generated
in standard single mode downconversion with sufficiently attenuated laser
light, so that the average photon numbers contributed by each source are approximately equal. By optimizing the ratio of coherent light and
downconverted light, it should then be possible to obtain $N$-photon
interference fringes with visibilities greater than $90\%$ for any
photon number $N$. The phase super-sensitivity of high photon number
path entangled states is thus much easier to obtain than previously
assumed.

Part of this work has been supported by the
Grant-in-Aid program of the Japanese Society for the
Promotion of Science and by the JST-CREST project on quantum
information processing.


\end{document}